\documentclass[11pt]{article}
\usepackage{amsfonts}
\usepackage{amsmath,amssymb}

\usepackage{amsfonts}
\usepackage{amsmath,amssymb}

\newcommand{\hko}{\hookrightarrow}

\newcommand{\n}{\nonumber\\}

\newcommand{\bu}{\bullet}

\newcommand{\cv}{\circ}

\newcommand{\bec}{\begin{center}}
\newcommand{\eec}{\end{center}}

\newcommand{\bea}{\begin{array}}
\newcommand{\ear}{\end{array}}

\newcommand{\bfr}{\begin{flushright}}

\newcommand{\efr}{\end{flushright}}
\newcommand{\noi}{\noindent}\newcommand{\Ra}{\rightarrow}

\newcommand{\cl}{{\mt{C}}\ell}

\newcommand{\RR}{\mathbb{R}}\newcommand{\op}{\oplus}

\newcommand{\la}{\Lambda}
\newcommand{\bege}{\begin{equation}}
\newcommand{\enge}{\end{equation}}
\newcommand{\w}{\wedge}

\newcommand{\ri}{\rightarrow}

\newcommand{\beq}{\begin{eqnarray}}\newcommand{\benu}{\begin{enumerate}}\newcommand{\enu}{\end{enumerate}}
\newcommand{\eeq}{\end{eqnarray}}
\newcommand{\mt}{\mathcal}

\newcommand{\ul}{\underline}

\newcommand{\non}{\nonumber\\}

\newcommand{\OO}{\mathbb{O}}

\newcommand{\bx}{\begin{pmatrix}}
\newcommand{\ex}{\end{pmatrix}}
\newcommand{\vcx}{\varepsilon}

\usepackage{indentfirst}  
\begin{document}
\title{Clifford algebra-parametrized octonions
 and generalizations}
\author{{\bf R. da Rocha}\thanks{DRCC - Instituto de F\'{\i}sica Gleb Wataghin, Unicamp, CP 6165, 13083-970, Campinas (SP), Brazil. 
E-mail: {\tt roldao@ifi.unicamp.br}. Supported by CAPES.}
\and
{\bf J. Vaz, Jr.}\thanks{Departamento de Matem\'atica Aplicada, IMECC, Unicamp, CP 6065, 13083-859, Campinas (SP), Brazil. 
 E-mail: 
{\tt vaz@ime.unicamp.br}}}

\date{}\maketitle

\abstract${}^{}$\begin{center}
\begin{minipage}{12cm}$ \;\;\;\;\;$ Introducing products between multivectors 
of $\cl_{0,7}$ (the Clifford algebra over the metric vector space $\RR^{0,7}$) and octonions, resulting in an octonion, and
leading to the non-associative standard  octonionic product  in a particular case, 
we generalize the octonionic $X$-product, associated with the transformation rules for bosonic and fermionic fields 
on the tangent bundle over the 7-sphere $S^7$, and the $XY$-product.
 This generalization is accomplished in the $u$- and $(u,v)$-products, 
 where $u,v\in\cl_{0,7}$ are fixed, but arbitrary. Moreover, 
we extend these original products in order to encompass the most general --- non-associative --- products 
 $(\RR\op\RR^{0,7})\times\cl_{0,7}\Ra \RR\op\RR^{0,7}$, $\cl_{0,7}\times(\RR\op\RR^{0,7})\Ra\RR\op\RR^{0,7}$ and $\cl_{0,7}\times\cl_{0,7}
\Ra \RR\op\RR^{0,7}$.   We also present the formalism necessary 
to construct Clifford algebra-parametrized octonions, which provides the structure  
to present the $\OO_{1,u}$ algebra.
Finally we introduce a method to construct $\OO$-algebras endowed with the $(u,v)$-product 
from $\OO$-algebras endowed with the $u$-product. These algebras are called $\OO$-like algebras and their octonionic units are
parametrized by arbitrary Clifford multivectors. 
 When $u$ is restricted to the underlying paravector space $\RR\op\RR^{0,7}\hko\cl_{0,7}$
of the octonion algebra $\OO$, these algebras 
are shown to be isomorphic. The products between Clifford multivectors and octonions, leading to an octonion, 
are shown to share graded-associative, supersymmetric properties. We also investigate the generalization of Moufang identities, 
for each one of the products introduced. 
\end{minipage}\end{center}
\medbreak\medbreak\medbreak\medbreak\noi
\bigskip\medbreak\noi
\medbreak\noi  MSC classification: 15A66, 17A35, 17C60, 81T60.

\newpage
\section{Introduction}

The $X$-product was originally introduced  in order to correctly define 
the transformation rules for bosonic [vector] and fermionic [spinor] fields 
on the tangent bundle over the 7-sphere $S^7$ \cite{ced}. This product is 
closely related to the parallel transport of sections of the tangent bundle, at $X\in S^7$, i.e., $X\in\OO$ such that $\bar{X}X = X\bar{X} = 1$. 
The $X$-product is also shown to be twice 
the parallelizing torsion \cite{mart}, given by the torsion tensor. This tensor is not constant due to the non-associativity of the octonion algebra 
$\OO$, 
 non-vanishing due to the non-commutativity of $\OO$, and in particular, it is used to investigate the 
 $S^7$ Ka$\check{{\rm c}}$-Moody algebra \cite{ced,kac}. The $X$-product  has also been used to obtain triality maps and $G_2$ actions \cite{dix,beng},
and it leads naturally to remarkable geometric and topological properties, for instance the Hopf fibrations   
$S^3\cdots S^7 \Ra S^4$ and $S^8\cdots S^{15} \Ra S^7$ \cite{dixpo, abg}, and twistor formalism in ten dimensions \cite{mart}.
The paramount importance of octonions in the search for unification is based, for instance, in the fact that by extending the 
division-algebra-valued superalgebras to octonions, in $D=11$ an octonionic generalized Poincar\'e superalgebra can be constructed, the 
so-called {octonionic $M$-algebra} that describes the octonionic $M$-theory \cite{10}, where the octonionic super-2-brane 
and  the octonionic super-5-brane sectors are shown to be equivalent \cite{top}. Also, there are other vast generalizations and applications
of the octonionic formalism \cite{baez,24} such as the classification of quaternionic and octonionic spinors \cite{carr} and the pseudo-octonionic formalism \cite{okubo}.

This paper is intended to
generalize the $X$- and $XY$-products, and to provide formal pre-requisites to define new Clifford algebra-parametrized octonions associated 
with the $(1,u)$-product, besides extending the original Moufang identities to the new products to be defined, when it is possible. 
In this sense, we produce a copy of $\OO$, but endowed with the $(1,u)$-product, instead of the standard octonionic product. 
Moreover we define and investigate
the fundamental properties of the $(1,u)$-, $u$- and $(u,v)$-products, $u,v\in\cl_{0,7}$, which naturally generalize the $X$- and $XY$-products, $X,Y\in\OO$
such that $\bar{X}X = X\bar{X} = 1 = \bar{Y}Y = Y\bar{Y}$, i.e., $X,Y\in S^7$. Octonionic products between Clifford algebra multivectors that results in
an octonion are also introduced, and the Moufang identities are shown to be generalizable for the $(1,u)$-product, while for other products it is 
shown not to be possible to extend the Moufang identities, even to new Moufang-like identities, obtained from the original ones using the automorphism
and anti-automorphisms of Clifford algebras.   
 This paper is organized as follows: Section 2 is devoted to present exterior and Clifford algebras and  
in Section 3 we review the 
fundamental properties of the octonionic algebra $\OO$, defined in terms of the Clifford algebra $\cl_{0,7}$ and its associated Clifford product
that defines the octonionic product \cite{loun}. The octonionic product is chosen to be defined in terms of the Clifford product 
in order that the arena of the whole formalism to be the Clifford algebra. In Section 4 the $X$-product and the $XY$-product \cite{abg} are generalized, encompassing 
`octonionic products' between arbitrary multivectors of $\cl_{0,7}$ and octonions.
After introducing products 
between Clifford multivectors and octonions, that results in an octonion, we illustrate the use of the formalism, giving 
some examples of useful computations. 
Section 5 is devoted to introduce octonionic, Clifford algebra-parametrized units, associated with the $(1,u)$-product and, using these Clifford algebraic-dependent octonionic units,
we assert the main results generalizing the formalism presented in \cite{abg}, this time encompassing the whole Clifford algebra $\cl_{0,7}$ instead of simply 
its paravector subspace $\RR\op\RR^{0,7}\hko\cl_{0,7}$. Although 
the product between $\cl_{0,7}$ and $\OO$ (that results in an octonion) is not associative, it is shown to be 
\emph{graded}-associative, if we apply the graded involution on elements of $\cl_{0,7}$. Also, the generalized Moufang identities, for the $(1,u)$-product,
 are  presented and discussed.
  In Section 6 an `octonionic product' between multivectors of $\cl_{0,7}$ is presented, which allows us to construct more 
general octonion algebras. It is shown that the Moufang identities are not valid for this product. 
Finally in Section 7 we construct $\OO$-like algebras with the 
octonionic product structure based on the just defined 
generalized octonionic products.

\section{Preliminaries}
Let $V$ be a finite $n$-dimensional real vector space. We consider the tensor algebra $\bigoplus_{i=0}^\infty T^i(V)$ from which we
restrict our attention to the space $\Lambda(V) = \bigoplus_{k=0}^n\Lambda^k(V)$ of multivectors over $V$. $\Lambda^k(V)$
denotes the space of the antisymmetric
 $k$-tensors, the $k$-forms.  Given $\psi\in\Lambda(V)$, $\tilde\psi$ denotes the \emph{reversion}, 
 an algebra antiautomorphism
 given by $\tilde{\psi} = (-1)^{[k/2]}\psi$ ([$k$] denotes the integer part of $k$). $\hat\psi$ denotes 
the \emph{main automorphism {\rm or} graded involution},  given by 
$\hat{\psi} = (-1)^k \psi$. The \emph{conjugation} is defined as the reversion followed by the main automorphism.
  If $V$ is endowed with a non-degenerate, symmetric, bilinear map $g: V\times V \rightarrow \RR$, it is 
possible to extend $g$ to $\la(V)$. Given $\psi=u_1\w\cdots\w u_k$ and $\phi=v_1\w\cdots\w v_l$, $u_i, v_j\in V$, one defines $g(\psi,\phi)
 = \det(g(u_i,v_j))$ if $k=l$ and $g(\psi,\phi)=0$ if $k\neq l$. Finally, the projection of a multivector $\psi= \psi_0 + \psi_1 + \cdots + \psi_n$,
 $\psi_k \in \la^k(V)$, on its $p$-vector part is given by $\langle\psi\rangle_p$ = $\psi_p$, while projection in its $p$ \emph{and} $q$ components is given by 
$\langle\psi\rangle_{p\op q} = \psi_p + \psi_q$. 
The Clifford product between $w\in V$ and $\psi\in\la(V)$ is given by $w\psi = w\w \psi + w\cdot \psi$.
 The Grassmann algebra $(\la(V),g)$ 
endowed with this product is denoted by $\cl(V,g)$ or $\cl_{p,q}$, the Clifford algebra associated with $V\simeq \RR^{p,q},\; p + q = n$.
\section{Octonions}
 The octonion algebra $\OO$ is defined as the paravector space $\RR\op\RR^{0,7}$ endowed with the product $\cv:(\RR\op\RR^{0,7})\times(\RR\op\RR^{0,7})\Ra\RR\op\RR^{0,7}$, 
denominated octonionic standard product. 
The identity $\{e_0 = 1\}\in\RR$ and an orthonormal basis $\{e_a\}_{a=1}^7$,
  in the underlying paravector space \cite{loun,bay2} $\RR\op\RR^{0,7}\hko\cl_{0,7}$ associated with $\OO$, generates the octonions. 
It is well known that the octonionic product can be constructed using the Clifford algebra $\cl_{0,7}$  as
\bege\label{eq1}
A\cv B = \langle AB(1 - \psi)\rangle_{0\oplus 1},\quad A,B\in\RR\op\RR^{0,7},
\enge\noi where $\psi = e_1e_2e_4 + e_2e_3e_5 + e_3e_4e_6 + e_4e_5e_7 + e_5e_6e_1 + e_6e_7e_2 + e_7e_1e_3 \in \Lambda^3(\RR^{0,7})\hko \cl_{0,7}$ and 
the juxtaposition denotes the Clifford product \cite{loun}. The idea of introducing the octonionic product from the Clifford one in this context is to 
present hereon our formalism using solely  Clifford algebras. Indeed, as $\OO$ is isomorphic to $\RR\op\RR^{0,7}$ \emph{as a vector space}, the octonionic
product fundamentally takes two arbitrary elements of the paravector space $\RR\op\RR^{0,7}$ --- which is itself endowed with the octonionic product ---
 resulting in another element of the paravector space. But looking to octonions inside the Clifford algebra arena
we can go beyond the paravector space and exploit the whole
Clifford algebra space, which is the way we use to generalize the
$X$- and $XY$-products.

It is now immediate, from Eq.(\ref{eq1}), to verify the usual rules between basis elements under the octonionic product\footnote{
Octonions are usually \emph{defined} by these rules, but here we point out the usefulness 
of considering these rules (Eq.(\ref{taboc})) as derived from Eq.(\ref{eq1}), using Clifford algebras.}:
\bege\label{taboc}
e_a\cv e_b = \vcx_{ab}^c e_c - \delta_{ab} \quad (a,b,c =1,\ldots,7), 
\enge\noi where we denote  $\vcx_{ab}^c = 1$
for the cyclic permutations ($abc$) = (124),(235),(346),(457),(561),(672) and (713).
Explicitly, the multiplication table is given by \cite{dix} 
\bec
\begin{tabular}{||r||r|r|r|r|r|r|r||}\hline\hline
 1&$e_1$&$e_2$&$e_3$&$e_4$&$e_5$&$e_6$&$e_7$\\\hline\hline
$e_1$&$-1$&$e_4$&$e_7$&$-e_2$&$e_6$&$-e_5$&$-e_3$\\\hline
$e_2$&$-e_4$&$-1$&$e_5$&$e_1$&$-e_3$&$e_7$&$-e_6$\\\hline
$e_3$&$-e_7$&$-e_5$&$-1$&$e_6$&$e_2$&$-e_4$&$e_1$\\\hline
$e_4$&$e_2$&$-e_1$&$-e_6$&$-1$&$e_7$&$e_3$&$-e_5$\\\hline
$e_5$&$-e_6$&$e_3$&$-e_2$&$-e_7$&$-1$&$e_1$&$e_4$\\\hline
$e_6$&$e_5$&$-e_7$&$e_4$&$-e_3$&$-e_1$&$-1$&$e_2$\\\hline
$e_7$&$e_3$&$e_6$&$-e_1$&$e_5$&$-e_4$&$-e_2$&$-1$\\\hline\hline\end{tabular}\label{tabela}
\eec\noi All the relations above can be expressed as $e_a\cv e_{a+1} = e_{a+3\mod 7}$.
Since we can consider the underlying vector space of $\OO$ as being $\RR\op\RR^{0,7}\hko\cl_{0,7}$, the Clifford conjugation of 
$X = x^0 + x^ae_a\in\OO$ is given by $\bar{X} = x^0 - x^ae_a$.

\section{The $u$-product and generalizations}
Hereon we suppose that $u\in\cl_{0,7}\equiv\cl(\RR^{0,7},g)$. 
Given fixed but arbitrary  $X,Y \in\RR\op\RR^{0,7}$ such that $X\bar{X} = \bar{X} X = 1 = \bar{Y}Y = Y\bar{Y}$ ($X,Y\in S^7$),
 the $X$-product is defined \cite{ced,mart,dix} by 
\bege
A\circ_X B:= (A\cv X)\circ(\bar{X} \cv B) = X\cv((\bar{X}\cv A)\cv B) = (A\cv (B\cv X))\cv \bar{X}.
\enge\noi 
 The $XY$-product is defined as:
\bege\label{xy}
A\circ_{X,Y} B:= (A\cv X)\cv(\bar{Y}\cv B).
\enge\noi In particular, the $(1,X)$-product is given by 
\bege\label{pro3}
A\circ_{1,X} B:= A\cv(\bar{X}\cv B).
\enge\noi  $X$ is the unit of $(1,X)$-product above, since $A\circ_{1,X} X = X\circ_{1,X} A = A$ \cite{ced,dix}.

We could propose a natural generalization of the $X$-product,
 introducing the $u$-product as:
\bege\label{pru}
A\circ_u B:= (Au)\cv(u^{-1}B).
\enge\noi But if the products $Au$ and $u^{-1}B$ are to be interpreted as Clifford products, the 
elements $u\in\cl_{0,7}$, such that $Au$ and $u^{-1}B$ are octonions, should be scalars. In this case $A \cv_u B \equiv A\cv B$
 and there would be nothing new to investigate.

In order that Eq.(\ref{pru}) to make sense, all  quantities between parenthesis must be octonions, 
and to avoid the trivial case (where $u$ must be a scalar), 
we have to define a product between octonions and Clifford multivectors that results in an octonion.

 Then, for  (homogeneous) multivectors  $u = u_1\ldots u_k\in \la^k(\RR^{0,7}) \hko \cl_{0,7}$, where $\{u_p\}_{p=1}^k \subset \RR^{0,7}$ ($k=1,\ldots,7)$ and 
$A\in\RR\op\RR^{0,7}$, we define the product $\bu_\llcorner$ as
\beq\label{def1}
\bu_\llcorner: (\RR\op\RR^{0,7})\times\la^k(\RR^{0,7}) &\Ra& \RR\op\RR^{0,7}\nonumber\\
(A,u) &\mapsto& A\bu_\llcorner u = ((\cdots(A\cv u_1)\cv u_2)\cv\cdots)\cv u_{k-1})\cv u_k.\label{def2}
\eeq\noi Note that the product $\bu_\llcorner$  is defined in such a way that, after the octonionic product between $A$ and $u_1$ is firstly performed,   the
 result is computed via the octonionic product, now,  with $u_2$. This process lasts up to the last octonionic product with $u_k$.
Parentheses in Eq.(\ref{def2}) emphasize the order of performance associated with the --- non-associative --- octonionic product, and
the symbol $\bu_\llcorner$ remembers us the octonion $A$ enters in the \emph{left} entry in the product in Eq.(\ref{def2}).

We also define the product $\bu_\lrcorner$ as 
\beq\bu_\lrcorner: \la^k(\RR^{0,7}) \times (\RR\op\RR^{0,7}) &\Ra& \RR\op\RR^{0,7}\nonumber\\
(u,A)&\mapsto& u\bu_\lrcorner A =  u_1\cv(\cdots\cv(u_{k-1}\cv(u_k\cv A))\cdots).\label{d3}
\eeq\noi Analogously, parentheses in Eq.(\ref{d3}) emphasize the order of performance associated with the octonionic product inside each parenthesis in Eq.(\ref{d3}), 
and the symbol $\bu_\lrcorner$ remembers us the element $A$ enters in the \emph{right} entry in the product in Eq.(\ref{d3}).

\medbreak
{\bf Remark 1}: It is clear that by extending the product in Eq.(\ref{def2}) to the scalars --- elements of $\la^0(\RR^{0,7})$ --- we have that $A \bu_\llcorner a = aA$, 
which denotes the trivial multiplication by scalars, where $a\in \RR = \la^0(\RR^{0,7})$.
Therefore now it is possible to extend by linearity the product $\bu_\llcorner$ to the whole exterior algebra $\la(\RR^{0,7}) = \oplus_{a=0}^7\la^a(\RR^{0,7})$ in such a way that the extended products are now denoted by 
\beq\label{def10}
\dot\bu_\llcorner: (\RR\op\RR^{0,7})\times\la(\RR^{0,7}) &\Ra& \RR\op\RR^{0,7}\\
\dot\bu_\lrcorner: \la(\RR^{0,7}) \times (\RR\op\RR^{0,7}) &\Ra& \RR\op\RR^{0,7}\label{def11}
\eeq
\medbreak
{\bf Remark 2}: In Remark 1 we have extended by linearity the products in Eqs.(\ref{def1}) and (\ref{d3}) from
$\la^k(\RR^{0,7})$ to the whole exterior algebra $\la(\RR^{0,7})$ in such a way that now $u\in\la(\RR^{0,7})$.
Therefore, when we restrict $u$ to the paravector space, i.e., $u\in\la^0(\RR^{0,7})\op\la^1(\RR^{0,7}) = \RR\op\RR^{0,7}\hko\la
(\RR^{0,7})$, then $u$ becomes an octonion, and 
 all the products $\dot\bu_\lrcorner$, $\dot\bu_\llcorner$, and the usual octonionic product $\circ$, are equivalent, since 
 all these products take in this case two elements of the paravector space $\RR\op\RR^{0,7}$ and maps them to another element of the paravector space, 
via the usual octonionic product defined by Eq.(\ref{eq1}).
 \medbreak

By abuse of notation we shall use hereon the symbol $\bu$ uniquely to denote both products $\dot\bu_\llcorner$ and $\dot\bu_\lrcorner$, 
in Eq.(\ref{def10}) and Eq.(\ref{def11}),
 and each one of the above-mentioned products are to be clearly implicit, as there exists an octonion in the left \emph{or} right entry 
of the product $\bu$.

Now, after the algebraic pre-requisites have been introduced, given $u\in\la(\RR^{0,7})$, the $u$-product is defined as
\beq
\circ_u: (\RR\op\RR^{0,7})\times(\RR\op\RR^{0,7})&\Ra&(\RR\op\RR^{0,7})\n
(A,B)&\mapsto& A\circ_u B:= (A\bu u)\cv(u^{-1}\bu B).
\eeq\noi It is worthwhile to note that 
\bege
A\cv_u B = (A\cv(B\bu u))\bu u^{-1} = u\bu((u^{-1}\bu A)\cv B).
\enge
\medbreak
\noi {\ul{Example 1}}: Let us calculate the product $e_1\cv_u e_4$, given $u = e_2e_7$:
\beq
e_1\circ_u e_4 &=& [e_1\bu (e_2e_7)]\cv [(e_2e_7)^{-1}\bu e_4]\non
&=& [(e_1\cv e_2)\cv e_7]\cv [-e_2\cv(e_7\cv e_4)]\non
&=& [e_4\cv e_7]\cv [-e_2\cv e_5]\non
&=& -e_5\cv (-e_3)\non
&=& -e_2.
\eeq \noi We note that $e_1\cv e_4 = -e_2$ too. We can prove, using Eq.(\ref{taboc}) and 
the property $\epsilon_{ijk}\epsilon_{lmk} = \delta_{il}\delta_{jm} - \delta_{im}\delta_{jl} + \vcx_{ijlm}$
that $A\cv_u B = A\cv B$, whenever $u$ is a homogeneous Clifford algebra element  of unit norm. 
When  $u$ is a paravector --- an element of $\RR\op\RR^{0,7}$ --- it is clear that the $u$-product is equivalent to the $X$-product.

In analogy to the $XY$-product and the ($1,X$)-product, just respectively defined by Eqs.(\ref{xy}) and (\ref{pro3}), 
it is also possible to define another product, the $(1,u)$-product, as 
\beq\label{jiri}
\cv_{1,u}:(\RR\op\RR^{0,7})\times(\RR\op\RR^{0,7}) &\Ra& \RR\op\RR^{0,7}\n
(A,B)&\mapsto& A\circ_{1,u} B:= A\cv(u^{-1}\bu B).
\eeq \noi
Finally, Eq.(\ref{xy}) can be generalized, given fixed $u,v\in\cl_{0,7}$, as:
\beq\label{uvb}
\cv_{u,v}:(\RR\op\RR^{0,7})\times(\RR\op\RR^{0,7}) &\Ra& \RR\op\RR^{0,7}\n
(A,B)&\mapsto& A\circ_{u,v} B:= (A\bu u)\cv(v^{-1}\bu B).
\eeq
\medbreak\noi 
{\ul{Example 2}}: Let us now calculate the product $e_1 \cv_{u,v} e_4$, where $u = e_4e_6e_7$ and $v=e_1 e_5$: 
\beq
e_1\circ_{u,v} e_4 &=& [e_1\bu (e_4e_6e_7)]\cv [(e_1e_5)^{-1}\bu e_4]\non
&=& [((e_1\cv e_4)\cv e_6)\cv e_7]\cv[-e_1\cv(e_5\cv e_4)]\non
&=& [(-e_2\cv e_6)\cv e_7]\cv [-e_1 \cv (-e_7)]\non
&=& (-e_7\cv e_7)\cv (-e_3)\non
&=& -e_3
\eeq

\section{$\OO$-units associated with the $(1,u)$-product}
We enunciate in what follows some results necessary to show that the octonion algebra $\OO$ is isomorphic to the algebra
 $\OO_{1,u} \equiv (\RR\op\RR^{0,7}, \cv_{1,u})$.
Lemmata 1, 2, 3, 4 and 5 are demonstrated by performing all 
the possible combinations of $\{e_a\}_{a=1}^7$ and then extending this process by linearity to the whole $\cl_{0,7}$. 
It is implicit that  $u\in\cl_{0,7}$ is \emph{not} a scalar, since in this case there would be nothing to prove. In what follows $A,B,C\in\OO$.
The next Lemma concerns the graded-associativity of the $\bu$-product. 

\medbreak
{\textbf{Lemma 1}}: ${}$ {\it The elements} 
$u\in\la^{2i}(\RR^{0,7})\hko\cl_{0,7}$ \emph{satisfy the relation} $(u\bu A)\cv B = u\bu(A\cv B)$,
 \emph{while the elements $u\in\la^{2i+1}(\RR^{0,7})\hko\cl_{0,7}$ satisfy the relation $(u\bu A)\cv B = -u\bu(A\cv B)$. These 
results can be expressed as}
$$ 
{}{(u\bu A)\cv B = \hat{u}\bu(A\cv B)} 
$$\noi \emph{Moreover, the relations}
$$ 
{}{A\cv(B\bu u) = (A\cv B)\bu \hat{u}} {}
$$\noi \emph{holds} for all $u\in\cl_{0,7}$.
 \medbreak
{\bf Remark 3}: Since we have just asserted that $A\bu B \equiv A\cv B$ when $A,B$ are octonions, the assertions of Lemma 1 are 
obviously equivalent to $(u\bu A)\bu B = \hat{u}\bu(A\bu B)$ and $A\bu(B\bu u) = (A\bu B)\bu \hat{u}$, from where it can be seen that we can
 correctly denominate this property as `graded-associativity'.
\medbreak

 {\textbf{Lemma 2}}: ${}$ {\it The elements $u\in \cl_{0,7}$ satisfy }
$$ 
{}{(u\bu A)\cv B = -(u\bu B)\cv A} {}
$$  \medbreak
 {\textbf{Lemma 3}}: ${}$ {\it The elements $u\in \cl_{0,7}\setminus\Lambda^6(\RR^{0,7})$ satisfy }
$$ 
{u\bu A =  A \bu \bar{u}}  {}
$$\noi \emph{In the particular case where $u = e_ae_be_ce_de_fe_g\in\Lambda^6(\RR^{0,7})$ and $A=e_h$, where none of the  subindices equals each other, 
the identity  $u\bu A = -1 = A\bu u$ holds}.  \medbreak
{\textbf{Lemma 4}}: ${}$  {\it The elements $u\in \cl_{0,7}$ satisfy the relation }
\bege
{}{u^{-1}\bu(u\bu A) = A = (A\bu u)\bu u^{-1}} {}
\enge
  \medbreak
\medbreak
 {\textbf{Lemma 5}}: ${}$  Given $u\in\cl_{0,7}$ it follows that
\bege
{}{u\bu (A\bu u) = (u\bu A)\bu u = u\bu A\bu u} {}
 \enge \medbreak \noi The  lemmata above allow to generalize the $\OO$-product,
also including `octonionic products' between octonions and multivectors of $\cl_{0,7}$. These lemmata provide the formal pre-requisite 
to construct the multiplicative table of the Clifford algebra-dependent units $\{E_A\}_{A=1}^7$, defined by
\begin{align}\label{baba}
E_1 &= u\bu e_1,\quad  E_2 = u\bu e_2,\quad E_3 = u\bu e_3,\quad E_4 = \hat{u}\bu e_4,\n
\quad E_5 &= u\bu e_5,\quad E_6 = \hat{u}\bu e_6\;\;\;{\rm and}\;\;\; E_7 = 
\hat{u}\bu e_1.\end{align}  

Doing the explicit verification on each one of the octonions defined by Eqs.(\ref{baba}), and subsequently extending by linearity in $\cl_{0,7}$, it follows the 
\medbreak
 {\textbf{Lemma 6}}: ${}$ \emph{The elements} $E_a$ \emph{anticommute under the }$(1,u)$-{\it product}, \emph{i.e}, 
$${}{E_a \cv_{1,u} E_b = - E_b \cv_{1,u} E_a} {}$$
  \medbreak
{\bf Proof}: When $\{a,b\} = \{1,2,3,5\}$, we see that
\beq
E_a \cv_{1,u} E_b &=& (u \bu e_a)\cv(u^{-1}\bu(u\bu e_b))\n
&=& (u \bu e_a)\cv e_b,\quad\text{using Lemma 4}\n
&=& -(u \bu e_b)\cv e_a,\quad\text{by Lemma 2}\n
&=& -(u \bu e_b)\cv(u^{-1}\bu(u\bu e_a))\n
&=& - E_b \cv_{1,u} E_a.
\eeq\noi The other cases, when $\{a,b\} = \{4,6,7\}$, are analogously demonstrated.
 \hspace{2,5cm}$\square$
\medbreak The  multiplication table of $\{E_a\}$ is inherited from  table (\ref{tabela}). 
 We express the table describing the $(1,u)$-product 
of the octonions $E_A$:
\bec
\begin{tabular}{||r||r|r|r|r|r|r|r||}\hline\hline
 1&$E_1$&$E_2$&$E_3$&$E_4$&$E_5$&$E_6$&$E_7$\\\hline\hline
$E_1$&$-1$&$E_4$&$E_7$&$-E_2$&$E_6$&$-E_5$&$-E_3$\\\hline
$E_2$&$-E_4$&$-1$&$E_5$&$E_1$&$-E_3$&$E_7$&$-E_6$\\\hline
$E_3$&$-E_7$&$-E_5$&$-1$&$E_6$&$E_2$&$-E_4$&$E_1$\\\hline
$E_4$&$E_2$&$-E_1$&$-E_6$&$-1$&$E_7$&$E_3$&$-E_5$\\\hline
$E_5$&$-E_6$&$E_3$&$-E_2$&$-E_7$&$-1$&$E_1$&$E_4$\\\hline
$E_6$&$E_5$&$-E_7$&$E_4$&$-E_3$&$-E_1$&$-1$&$E_2$\\\hline
$E_7$&$E_3$&$E_6$&$-E_1$&$E_5$&$-E_4$&$-E_2$&$-1$\\\hline\hline\end{tabular}\label{tabelaw}
\eec\noi
With this table we immediately see that $\{E_A\}_{A=1}^7$ are the octonion units associated with the $(1,u)$-product. 
We exhibit below the computations related to the first 
line of the table above. The other lines follow in an analogous way.
\beq 
E_1 \cv_{1,u} E_2 &=& (u\bu e_1)\cv (u^{-1}\bu(u\bu e_2))\n
&=& (u\bu e_1)\cv e_2, \quad \text {using Lemma 4}\n
&=& \hat{u}\bu(e_1\cv e_2),\quad \text {by Lemma 1}\n
&=& \hat{u}\bu e_6\n
&=& E_6\eeq
We also have
\beq
E_1 \cv_{1,u} E_3 &=& (u\bu e_1)\cv (u^{-1}\bu(u\bu e_3)) = (u\bu e_1)\cv e_3\n
&=& \hat{u}\bu(e_1\cv e_3) = \hat{u}\bu e_4\n
&=& E_4,\eeq
\beq
E_1 \cv_{1,u} E_4 = - E_4 \cv_{1,u} E_1 &=& -(\hat{u}\bu e_4)\cv (u^{-1}\bu(u\bu e_1))\n
&=& -(\hat{u}\bu e_4)\cv e_1 = -{u}\bu (e_4\cv e_1) = -u\bu e_3\n
&=& -E_3,\eeq
\beq
E_1 \cv_{1,u} E_5 &=& (u\bu e_1)\cv (u^{-1}\bu(u\bu e_5))\n
&=& (u\bu e_1)\cv e_5 = \hat{u}\bu(e_1\cv e_5)= \hat{u}\bu e_7\n
&=& E_7,\eeq
\noi  
\beq
E_1 \cv_{1,u} E_6 = - E_6 \cv_{1,u} E_1 &=& -(\hat{u}\bu e_6)\cv (u^{-1}\bu(u\bu e_1))\n
&=& -(\hat{u}\bu e_6)\cv e_1 = -{u}\bu (e_6\cv e_1)= -u\bu e_2\n
&=& -E_2.\eeq
\noi 
\medbreak
{\bf Remark 4}: The Moufang identities \cite{dix,23} 
 \bege\label{mou0}
(A\cv B)\cv(C\cv A) = A\cv (B\cv C)\cv A,
\enge\noi 
\bege\label{mou1}
(A\cv B\cv A)\cv C = A\cv(B\cv(A\cv C)),
\enge
\bege\label{mou2}
(A\cv B)\cv (C\cv A) = A\cv(C\cv B)\cv A,
\enge
\bege\label{mou3}
C\cv(A\cv B\cv A) = ((C\cv A)\cv B)\cv A,
\enge\noi $ A,B,C\in\OO$ can be immediately generalized for the ($1,u$)-product, using table (\ref{tabelaw}), as
 \bege
(A\cv_{1,u} B)\cv_{1,u}(C\cv_{1,u} A) = A\cv_{1,u} (B\cv_{1,u} C)\cv_{1,u} A,
\enge\noi 
\bege
(A\cv_{1,u} B\cv_{1,u} A)\cv_{1,u} C = A\cv_{1,u}(B\cv_{1,u}(A\cv_{1,u} C)),
\enge
\bege
(A\cv_{1,u} B)\cv_{1,u} (C\cv_{1,u} A) = A\cv_{1,u}(C\cv_{1,u} B)\cv_{1,u} A,
\enge
\bege
C\cv_{1,u}(A\cv_{1,u} B\cv_{1,u} A) = ((C\cv_{1,u} A)\cv_{1,u} B)\cv_{1,u} A,
\enge\noi In the case of the  products
\bege\dot\bu_\llcorner: (\RR\op\RR^{0,7})\times\la(\RR^{0,7}) \Ra \RR\op\RR^{0,7}\;\;\;{\rm  and}\;\;\;  \dot\bu_\lrcorner: \la(\RR^{0,7})\times (\RR\op\RR^{0,7})
\Ra \RR\op\RR^{0,7},\enge\noi it can be shown 
that 
uniquely using the Clifford conjugation and the graded involution it is not possible to get some kind of Moufang identities. 
 The following counterexamples show why it is not possible.
\medbreak {\ul{Example 3}}: One of the Moufang identities for octonions is expressed as  
\bege(A\cv B)\cv(C\cv A) = A\cv (B\cv C)\cv A,\quad A,B,C\in\OO.\enge\noi  
Suppose that an immediate generalization for such an identity is the expression
\bege\label{relg}
(u\bu A)\bu(B\bu u) = u\bu(A\bu B)\bu u,\quad u\in\cl_{0,7},
\enge\noi or $(u\bu A)\bu(B\bu u) = \hat{u}\bu(A\bu B)\bu u$, 
or $(u\bu A)\bu(B\bu u) = \bar{u}\bu(A\bu B)\bu u$, or the product given by Eq.(\ref{relg}), 
for any combination of the graded involution and/or Clifford conjugation acting on $u$. 
In order to the expressions become more clear, we denote Eq.(\ref{relg}) as
\bege\label{relga}
(u\bu A)\cv(B\bu u) = u\bu(A\cv B)\bu u,\quad u\in\cl_{0,7},
\enge\noi since the product $\bu$ between paravectors --- elements of $\RR\op\RR^{0,7}$ --- is equal ot the product $\cv$.
Let $u = e_6e_7e_1e_3$, $A = e_2$ and $B = e_5$. On the one hand, 
\bege(e_6e_7e_1e_3\bu e_2)\cv(e_5\bu e_6e_7e_1e_3) = -e_4,
\enge\noi
 and \bege e_6e_7e_1e_3\bu (e_2\cv e_5) \bu e_6e_7e_1e_3 = -e_4.
\enge\noi
 On the other hand, if we take  $u = e_1e_2e_3e_6$, $A = e_4$ and $B = e_7$, we have that: 
\bege (e_1e_2e_3e_6\bu e_4)\cv(e_7\bu e_1e_2e_3e_6) = e_6,
\enge\noi
 while 
\bege e_1e_2e_3e_6\bu (e_4\cv e_7) \bu e_1e_2e_3e_6 = -e_6.\enge\noi
 We then realize that for distinct elements  $u\in\Lambda^4(\RR^{0,7})$, we have  
$(u\bu A)\cv(B\bu u) = u\bu(A\cv B)\bu u$, and 
$(u\bu A)\cv(B\bu u) = -u\bu(A\cv B)\bu u$. 
These last two relations cannot be mutually satisfied by two elements in $\cl_{0,7}$ presenting the same degree. 
Analogous counterexamples can be shown for the product given by Eq.(\ref{relg}) with any combination of the graded involution and/or Clifford conjugation acting on $u$. 
It is also possible to show that the other Moufang identities given by Eqs.(\ref{mou1}, \ref{mou2}, \ref{mou3}) cannot 
be generalized for the products $\bu: (\RR\op\RR^{0,7})\times\la(\RR^{0,7}) \Ra \RR\op\RR^{0,7}$ and 
 $\bu: \la(\RR^{0,7})\times (\RR\op\RR^{0,7}) \Ra \RR\op\RR^{0,7}$, uniquely 
using the Clifford conjugation and the graded involution.  

\section{Octonionic product between Clifford  multivectors and generalizations}

Given vectors $\{u_p\}_{p=1}^k\subset \RR^{0,7}$ and $\{v_q\}_{q=1}^k\subset \RR^{0,7}$, $1\leq k\leq 7$, and
elements $u = u_1\ldots u_k, v = v_1\ldots v_k \in \cl_{0,7}$, we
 define the ``octonionic product'' between elements of the Clifford algebra $\cl_{0,7}$ as
\beq
\odot_\llcorner: \cl_{0,7} \times \cl_{0,7} &\ri& \RR\op\RR^{0,7}\nonumber\\ 
(u,v)&\mapsto&{u\odot_\llcorner v:=  u_1\cv(u_2\cv(\cdots\cv(u_k\bu v)\cdots))}\label{1087}
\eeq\noi Parentheses in Eq.(\ref{1087}) emphasize the order of performance associated with the non-associative octonionic product. After calculating 
the product $u_k\bu v$, the result is computed via the octonionic product with $u_{k-1}$, and then successively until the last reminiscent result, which is computed
with $u_1$ through the usual octonionic product.
It can be easily seen that, when we restrict elements of $\cl_{0,7}$ to the paravector space $\RR\op\RR^{0,7}$, then $A\odot_\llcorner B \equiv A \cv B$, where 
$A, B\in\RR\op\RR^{0,7}$. 
\medbreak
\noi {\ul{Example 4}}: Let us compute the product $e_1e_2\odot_\llcorner e_3e_4$:
\beq
e_1e_2\odot_\llcorner e_3e_4 &=& e_1\cv(e_2\bu(e_3e_4)) = e_1\cv(e_5\cv e_4) = e_1\cv (-e_7) \n
&=&  e_3.
\eeq
\medbreak
\noi We also define the product
\beq\label{lr3}
 \odot_\lrcorner: \cl_{0,7} \times \cl_{0,7} &\ri& \RR\op\RR^{0,7}\nonumber\\
(u,v)&\mapsto&{u\odot_\lrcorner v:=  ((\cdots\cv(u\bu v_1)\cv v_2)\cv\cdots)\cv v_k}
\eeq\noi It is immediate to see that $A\odot_\lrcorner B = A \cv B$, whenever $A,B\in\RR\op\RR^{0,7}$, since we have just seen 
that $A\bu B$ = $A\circ B$, by Eqs.(\ref{def2,d3}). 

It is now immediate to see that the Moufang identities (\ref{mou0}, \ref{mou1}, \ref{mou2}, \ref{mou3}) are not valid to the products $\odot_\lrcorner$ 
and $\odot_\llcorner$. 
\medbreak
\noi {\ul{Example 5}}: Let us see whether the Moufang identity given by Eq.(\ref{mou0}) can be generalized to the $\odot_\llcorner$ product. 
On the one hand let us compute the product $e_7e_3\odot_\llcorner(e_5e_4\odot_\llcorner e_1e_6) \odot_\llcorner e_7e_3$:
\beq
e_7e_3\odot_\llcorner(e_5e_4\odot_\llcorner e_1e_6) \odot_\llcorner e_7e_3 &=& e_7e_3\odot_\llcorner(e_5 \cv (e_4\bu e_1e_6)) \odot_\llcorner e_7e_3 \n
&=& e_7e_3\odot_\llcorner(e_5 \cv (e_2\cv e_6)) \odot_\llcorner e_7e_3\n
&=& e_7e_3\odot_\llcorner(e_5 \cv e_7) \odot_\llcorner e_7e_3\n
&=& e_7e_3\odot_\llcorner e_4  \odot_\llcorner e_7e_3\n
&=& (e_7\cv e_6) \bu e_7e_3\n
&=& -e_2\bu e_7e_3\n
&=& e_6\cv e_3\n
&=&e_4.
\eeq
On the other hand we have:
\beq (e_7e_3\odot_\llcorner e_5e_4)\odot_\llcorner (e_1e_6 \odot_\llcorner e_7e_3) &=& (e_7 \cv (e_2\cv e_4))\cv (e_1\cv(e_2\cv e_3))\n
&=& (e_7\cv e_1) \cv (e_1\cv e_5)\n
&=& e_3\cv e_6\n
&=& -e_4.
\eeq\noi \medbreak Then a counterexample for the validity of the Moufang identity Eq.(\ref{mou0}), for the $\odot_\llcorner$, was exhibited, and
 similarly it is easy to see that the other identities given by Eqs.(\ref{mou1}, \ref{mou2}, \ref{mou3}) are not generalizable too. 

Definition (\ref{lr3}) allows us to see that the $(1,u)$-product can also be generalized in order to encompass elements $u\in\cl_{0,7}$ in the first or in the second entry,
as follows (by abuse of notation we denote  distinct products $\cv_{1,u}:(\RR\op\RR^{0,7})\times(\RR\op\RR^{0,7}) \Ra \RR\op\RR^{0,7}, \;\cv_{1,u}:\cl_{0,7}\times(\RR\op\RR^{0,7}) \Ra \RR\op\RR^{0,7}$, 
\; $\cv_{1,u}:(\RR\op\RR^{0,7})\times\cl_{0,7} \Ra \RR\op\RR^{0,7}$ and $\cv_{1,u}:\cl_{0,7}\times\cl_{0,7} \Ra \RR\op\RR^{0,7}$ by the same symbol $\cv_{1,u}$):
\beq\label{jiripo}
\cv_{1,u}:\cl_{0,7}\times(\RR\op\RR^{0,7}) &\Ra& \RR\op\RR^{0,7}\n
(v,A)&\mapsto& v\circ_{1,u} A:= v\bu(u^{-1}\bu A).
\eeq \noi Hereon we can opt to use $\odot_\lrcorner$ or $\odot_\llcorner$ in the definitions below, and therefore we adopt the symbol $\odot$ to denote any one of them.
 Obviously if one of the two
products $\odot_\lrcorner$ and $\odot_\llcorner$ is to be chosen, the respective  
 choice must also be made in all the following definitions. Now we define
\beq\label{jiki}
\cv_{1,u}:(\RR\op\RR^{0,7})\times\cl_{0,7} &\Ra& \RR\op\RR^{0,7}\n
(A,v)&\mapsto& A\circ_{1,u} v:= A\cv(u^{-1}\odot v),
\eeq \noi
 It is immediate that by Eq.(\ref{jiripo}) we have $A\cv_{1,u} u = A$, so the element $u$ is the right unit of the product defined in Eq.(\ref{jiki}).
Using Lemma 4 and Eq.(\ref{jiripo}) we can also prove that $u\cv_{1,u} A = A$, so that $u$ is also the left unit related 
to the product defined in Eq.(\ref{jiki}), and we conclude that the unit associated with the $(1,u)$-product is $u\in\cl_{0,7}$.
The last $\cv_{1,u}$ extension, given $z\in\cl_{0,7}$ fixed but arbitrary, is defined by
\beq\label{jiba}
\cv_{1,u}:\cl_{0,7}\times\cl_{0,7} &\Ra& \RR\op\RR^{0,7}\n
(v,z)&\mapsto& v\circ_{1,u} z:= v\bu(u^{-1}\odot z),
\eeq \noi

Given $z,t\in\cl_{0,7}$, the products given in Section 4 are immediately extended\footnote{We analogously denote the products below 
by the same symbols, by abuse of notation.} if we define:
\beq
\circ_u: \cl_{0,7}\times(\RR\op\RR^{0,7})&\Ra&\RR\op\RR^{0,7}\n
(v,A)&\mapsto& v\circ_u A:= (v\odot u)\cv(u^{-1}\bu A),
\eeq\noi 
\beq
\circ_u: (\RR\op\RR^{0,7})\times\cl_{0,7}&\Ra&\RR\op\RR^{0,7}\n
(A,v)&\mapsto& A\circ_u v:= (A\bu u)\cv(u^{-1}\odot v),
\eeq\noi 
\beq
\circ_u: \cl_{0,7}\times\cl_{0,7}&\Ra&\RR\op\RR^{0,7}\n
(v,z)&\mapsto& v\circ_u z:= (v\odot u)\cv(u^{-1}\odot z).
\eeq\noi Finally the $(u,v)$-product defined by Eq.(\ref{uvb}) can also be extended as: 
\beq
\cv_{u,v}:\cl_{0,7}\times(\RR\op\RR^{0,7}) &\Ra& \RR\op\RR^{0,7}\n
(z,A)&\mapsto& z\circ_{u,v} A:= (z\odot u)\cv(v^{-1}\bu A),
\eeq
\beq
\circ_{u,v}: (\RR\op\RR^{0,7})\times\cl_{0,7}&\Ra&\RR\op\RR^{0,7}\n
(A,z)&\mapsto& A\circ_{u,v} z:= (A\bu u)\cv(v^{-1}\odot z),
\eeq\noi 
\beq
\circ_{u,v}: \cl_{0,7}\times\cl_{0,7}&\Ra&\RR\op\RR^{0,7}\n
(t,z)&\mapsto& t\circ_{u,v} z:= (t\odot u)\cv(v^{-1}\odot z).
\eeq\noi

\section{Generalizations and equivalence between $\OO$-like algebras}

From table (\ref{tabelaw}), we see that $\OO\simeq \OO_{1,u}$. Now it is shown how, from a copy of $\OO_u$, we can obtain for instance 
the algebra $\OO_{u,C}:=(\RR\op\RR^{0,7}, \cv_{u,C})$, $u\in \cl_{0,7}$ and $C$ is an octonion. The process is accomplished by taking in the algebra $\OO_u$ = ($\OO, \cv_{u}$), 
products-$(1,v)$, i.e., given $A,B\in\RR\op\RR^{0,7}$, we calculate the  product:
\beq\label{acb}
A\cv_u(v^{-1}\cv_u B) &=& (A\bu u)\cv \{u^{-1}\bu[(v^{-1}\odot u)\cv(u^{-1}\bu B)]\}\n
&=& (A\bu u)\cv \{u^{-1}\bu[u\bu((u^{-1}\odot v^{-1})\cv B)]\}\n
&=&  (A\bu u)\cv [(u^{-1}\odot v^{-1})\cv B)]\n
&=& (A\bu u)\cv (C\cv B)\n
&=& A \cv_{u,C} B,
\eeq\noi where we have defined \bege\label{ofghj} C = u^{-1}\odot v^{-1}\in \RR\op\RR^{0,7}.\enge 
The algebra $\OO_u$ then originates $\OO_{u,C}$. 

In an analogous way the algebra $\OO_u$ gives rise to the algebra $\OO_{B\bu u}$, by the
 \medbreak
 {\textbf{Theorem 1}} ${}$ $(A \cv_u B) \cv_u ({B}^{-1} \cv_u C) = \pm A \cv_{B\bu u} C$, $\;\;\;\forall A,B,C\in \RR\op\RR^{0,7}, u\in\cl_{0,7}$ ${}$
\medbreak
{\bf Proof}: 
\beq
(A \cv_u B) \cv_u (B^{-1} \cv_u C) &=& \{[(A\bu u)\cv (u^{-1}\bu B)]\bu u\}\cv\{u^{-1}\bu[({B}^{-1}\bu u)\cv(u^{-1}\bu C)]\}\n
&=& \{[(A\cv(B\bu u))\bu u^{-1}]\bu u\}\cv\{u^{-1}\bu[u\bu((u^{-1}\bu {B}^{-1})\cv C)]\}\n
&=& [A\cv (B\bu u)]\cv[(u^{-1}\bu {B}^{-1})\cv C]\n
&=& \pm [A\cv (B\bu u)]\cv[(B\bu u)^{-1}\cv C]\n
&=& \pm A \cv_{B\bu u} C.
\eeq
 \noi If $B\in\RR\op\RR^{0,7}$ is such that $B \bar{B} = \bar{B}B = 1$, i.e., $B\in S^7$, then Theorem 1 is enunciated as 
\bege{} (A \cv_u B) \cv_u (\bar{B} \cv_u C) =  \pm A \cv_{B\bu u} C, \quad\forall A,B,C\in \RR\op\RR^{0,7}, u\in\cl_{0,7}{}
\enge

\section{Concluding Remarks}

We generalized the $X$- and $XY$-products, introducing the products $\cv_{u}:(\RR\op\RR^{0,7})\times(\RR\op\RR^{0,7})
\Ra\RR\op\RR^{0,7}$ and $\cv_{u,v}:(\RR\op\RR^{0,7})\times(\RR\op\RR^{0,7})\Ra\RR\op\RR^{0,7}$,
 where $u,v\in\cl_{0,7}$ are chosen to be fixed but arbitrary. After the formal definitions we extended these products in order to encompass 
the products $(\RR\op\RR^{0,7})\times\cl_{0,7}\Ra \RR\op\RR^{0,7}$, $\cl_{0,7}\times(\RR\op\RR^{0,7})\Ra\RR\op\RR^{0,7}$ and $\cl_{0,7}\times\cl_{0,7}\Ra \RR\op\RR^{0,7}$. 
We also furnish the mathematical requirements to introduce  Clifford algebra-parametrized octonionic units associated 
with the $(1,u)$-product and, consequently, to produce a copy of $\OO$, but now endowed with the $(1,u)$-product. The Moufang identities are 
immediately obtained for such new octonionic units.  Also, 
the algebra $\OO_{u}$ and $\OO_{B\bu u}$ are shown to be equivalent, since the former originates the latter from iterated $u$-products. 
The products $\bu:\cl_{0,7}\times(\RR\op\RR^{0,7})\Ra\RR\op\RR^{0,7}$ and $\bu:(\RR\op\RR^{0,7})\times\cl_{0,7}\Ra\RR\op\RR^{0,7}$ are 
{graded}-associative, indicating the rise of a supersymmetric structure out of these products.
Some more general algebras can be constructed from simpler ones, as the explicit construction given by Eq.(\ref{acb}). 
Although the products  $\bu_\lrcorner:\la(\RR^{0,7})\times(\RR\op\RR^{0,7})
\Ra\RR\op\RR^{0,7}$ and $\bu_\llcorner:(\RR\op\RR^{0,7})\times\la(\RR^{0,7})\Ra\RR\op\RR^{0,7}$
are equivalent, in a sense, to the left and right actions presented in \cite{dix}, this new approach allows us to generalize these actions in order to enclose 
all the products defined in this paper. Moreover, the present approach permits the graded-associativity shown in Lemma 1 to become transparent, and
to use a simpler notation. Finally, by considering octonions inside the Clifford algebra arena
we can go beyond the paravector space $\RR\op\RR^{0,7}$ and approach 
Clifford algebra $\cl_{0,7}$ space, which is the most natural formalism we use to generalize the
$X$ and $XY$-products. Explicit applications of the present formalism in some physical theories are to be presented in a forthcoming paper \cite{daro1}.

\section*{Acknowledgements}
The authors thank to Journal of Algebra Referee
for elucidating and enlightening viewpoints, and also Dr. R. A. Mosna for useful discussions.

\end{document}